\documentclass[aps,pra,reprint,superscriptaddress]{revtex4-2}
\usepackage{graphicx} 
\usepackage{physics}
\usepackage{siunitx}
\usepackage{amsfonts,amssymb,amsmath}
\usepackage{xcolor}
\usepackage{hyperref}
\usepackage{bbold}
\usepackage{mathtools}

\newcommand{\nuc}{\textsuperscript{13}C }

\newcommand{\uproman}[1]{\uppercase\expandafter{\romannumeral#1}}

\begin{document}
\title{Control Protocols for Entangling Gates for Group-IV Color-Centers in Diamond} 

\author{Jurek Frey}
\affiliation{Peter Gr\"unberg Institute-Quantum Computing Analytics (PGI-12), Forschungszentrum J\"ulich GmbH, D-52425 J\"ulich, Germany}
\affiliation{Theoretical Physics, Saarland University, 66123 Saarbr\"ucken, Germany}
\author{Frank K. Wilhelm}
\affiliation{Peter Gr\"unberg Institute-Quantum Computing Analytics (PGI-12), Forschungszentrum J\"ulich GmbH, D-52425 J\"ulich, Germany}
\affiliation{Theoretical Physics, Saarland University, 66123 Saarbr\"ucken, Germany}
\date{\today}
\author{Matthias M. M\"uller}
\affiliation{Peter Gr\"unberg Institute-Quantum Control (PGI-8), Forschungszentrum J\"ulich GmbH, D-52425 J\"ulich, Germany}
\thanks{Corresponding author: \href{mailto:ma.mueller@fz-juelich.de}{ma.mueller@fz-juelich.de}}
\begin{abstract}
Accurately controlling entangling gates remains a major challenge for quantum technology applications with solid-state spin qubits.
Here, we study a group-IV color-center with a strongly-coupled nuclear spin and approach the problem from a quantum control perspective.
We show that there are three different types of entangling gates where the entanglement is mediated by the parallel hyperfine-coupling component, the orthogonal one or both.
We derive the respective quantum speed limits (QSL) and show by means of dynamical decoupling, resonant driving of single- and double-quantum transitions, quantum optimal control and algebraic gate decomposition how these gates can be realized. 
We finally discuss the experimental applicability.
\end{abstract}
%
\maketitle
%
\section{Introduction}
\label{sec:introduction}
Reliable quantum applications such as distributed computing~\cite{wei2025universal}, entanglement distribution~\cite{oleynik2025entanglement} and non-local quantum sensing~\cite{stas2025entanglement} rely heavily on the quality and operational capabilities of the underlying building blocks in form of quantum-network nodes~\cite{halder2024optimal, guo2020distributed,majumder2026}. 
In particular, color centers in diamond, including the widely studied nitrogen-vacancy 
(NV) center, provide a versatile framework for realizing the necessary spin-photon 
interfaces \cite{schirhagl2014nitrogen, neumann2012towards}.
However, many control protocols for the NV center either rely on the intrinsic (nitrogen) host spin which is not available for group-IV centers~\cite{harris2023hyperfine} or on the asymmetry in the hyperfine interaction due to the $m_s=0$ electron spin state, while the symmetric hyperfine interaction for group-IV centers makes these protocols more challenging~\cite{beukers2025control}.
In this class, germanium vacancy (GeV) centers in diamond have emerged as interesting candidates for realizing such quantum nodes due to their superior optical transition properties, long-lived nuclear memory and strong hyperfine coupling~\cite{harris2024coherence, grimm2025coherent, patel2024single, knaut2024entanglement, ruf2021quantum}.
\\
The split-vacancy GeV defect, formed by a substitutional germanium atom and an adjacent vacancy, hosts an electronic spin-$\frac{1}{2}$ that couples to nearby \nuc carbon nuclear spins~\cite{harris2024coherence, grimm2025coherent}.
The surrounding \nuc nuclear spins enable the storage and entanglement of quantum information by forming a long-lived spin register~\cite{grimm2025coherent, taminiau2014universal, abobeih2018one} providing an excellent resource for error-correction protocols.
In these systems fast and high-fidelity two-qubit gate operations are a bottleneck for network-node operation rates~\cite{pompili21realization, childress2013diamond, kalb17entanglement}.
To approach this bottleneck several strategies have been explored for effective control of such registers.
The nuclear spins can for example be addressed directly using radio-frequency (RF) fields~\cite{grimm2025coherent}.
Additionally, control over the nuclear spins and over the interaction terms can be mediated via the electron spin through dynamical decoupling sequences or driving double-quantum transitions (DQT)~\cite{patel2024single, taminiau2014universal, bradley2019ten, takou2024generation}. 
Quantum optimal control (QOC) techniques have demonstrated the full controllability of electron-nuclear systems via microwave (MW) control~\cite{rembold2020,chou2015optimal,Yuan2015,wang2025, frey2026optimaltwoqubitgatesgroupiv}.

Once a color center is selected, the positions and therefore the interaction strength of the surrounding \nuc nuclear spins is inherently random and the resulting hyperfine coupling terms $A_{zx}$ and $A_{zz}$ set the available interaction channels~\cite{hanson2008coherent, taminiau2014universal, maze2008electron}. 
As we show in this study, the strength of the interaction terms plays an important role in which protocol is most practical and how fast the gates are implementable.
We note that the gate duration directly impacts the decoherence-limited fidelity and therefore quick operation improves the performance~\cite{nielsen2010quantum}.

Here, we systematically analyze existing two-qubit gate protocols such as driving double quantum transitions~\cite{travesedo2025all} which we fine-tune with the help of QOC, dynamical-decoupling-based gate design~\cite{taminiau2014universal, zhao2012decoherence,takou2024generation} where we present an alternative QOC solution, and we apply a method for algebraic gate synthesis~\cite{kraus2001optimal, khaneja2001cartan, makhlin2002nonlocal, watts2015optimizing,wang2025,baran2026} to the spin-1/2 color center coupled to a single \nuc nuclear spin.
Furthermore, we compare these protocols with regard to speed, experimental feasibility and sensitivity to parameter variations, by identifying  the distinct operating regimes as a function of the underlying parameters $A_{zz}, A_{zx}$ and $\omega_I$. 
For realistic benchmarking of the system all simulations use the experimentally extracted system parameters reported in~\cite{grimm2025coherent,Gundlapalli2025,frey2026optimaltwoqubitgatesgroupiv}, but the main findings are applicable to all group-IV centers and beyond.
The system analysis is intentionally performed under coherent Hamiltonian evolution focusing on theoretical insights on the energy structure rather than the noise mitigation strategies that were analyzed in Ref.~\cite{frey2026optimaltwoqubitgatesgroupiv}.
In particular, we want to investigate the quantum speed limit (QSL) for these gate time, i.e., the fastest possible gate time that arises from the bounded-operator structure of the driven system~\cite{Margolus1998,Caneva2009,Deffner2013,Deffner2017}.

The paper is organized in the following way: 
In ~\autoref{sec:background} we introduce the system Hamiltonian, the frame we want to operate in and derive an effective Hamiltonian.
In ~\autoref{sec:Two-Qubit-Protocols} we present our results for the three gate protocols and provide the analytic conditions for implementing them. 
Subsequently, in ~\autoref{sec:discussion} we compare the methods in terms of fidelity, gate time and experimental feasibility and conclude in ~\autoref{sec:conclusion} with discussing  the implications for magnetic-field choice and provide practical gate-set design guidelines.  
\section{Background and Theory}
\label{sec:background}
\subsection{Two-Qubit System}
\begin{figure}
    \centering
    \includegraphics{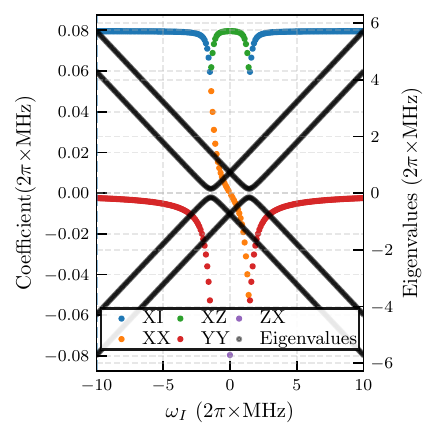}
    \caption{Shown in color is the Pauli decomposition of the Hamiltonian from Eq.~\eqref{eq:eigenframe}, where the order of the eigenvectors was the same as in the original Hamiltonian's diagonal. The gray dots show the eigenvalues of the Hamiltonian, two avoided crossings and a real degeneracy at $\omega_I = 0$ are visible. 
    }
    \label{fig:pauli_decomp_ham}
\end{figure}

The electronic spin associated with the GeV center couples to nearby \nuc nuclear spins via a combination of dipole-dipole and Fermi contact hyperfine interactions~\cite{harris2024coherence, grimm2025coherent, karim2023all}. 
The strength and anisotropy of this coupling depend on the relative position and orientation of the nuclear spin to the magnetic field~\cite{harris2024coherence, karim2023all}. 
For a single, strongly coupled nuclear spin, the effective Hamiltonian can, after the secular approximation (with $\hbar=1$), be written as 
\begin{equation}
    H_0 = -\omega_e S_z - \omega_I I_z + A_{zz} S_z I_z + A_{zx} S_z I_x, 
\end{equation}
where $\omega_e = \gamma_e B$ and $\omega_I = \gamma_I B$ are the electronic and nuclear Zeeman splittings at the magnetic field $B$, while $\gamma_e, \gamma_n$ are the effective gyromagnetic ratios of the electronic and nuclear spin, and $A_{zz}$ ($A_{zx}$) denote the longitudinal (transverse) hyperfine couplings.
$S_k$ are the spin operators acting on the electron spin and $I_k$ are the spin operators on the nuclear spin.
While $A_{zx}$ and $A_{zz}$ are fixed by the geometry of the Germanium vacancy center and its surrounding (and we choose $A_{zz} = 2\pi \times \SI{2.86234}{\MHz}$, and $A_{zx} = 2\pi \times \SI{0.60281}{\MHz}$ corresponding to the experimental setup studied in~\cite{grimm2025coherent,Gundlapalli2025, frey2026optimaltwoqubitgatesgroupiv}), the Larmor frequency can be tuned by changing the applied magnetic field. Throughout this work we choose $\omega_I \in 2\pi \times [-15,15]\,\SI{}{\MHz}$ and specify the value where needed.
Control of the electron spin is achieved by a MW drive,
\begin{equation}
    H_c = 2\Omega(t) S_x \cos(\omega_c t - \varphi),
\end{equation}
where $\Omega$ is the time-dependent Rabi frequency and $\omega_c$ the carrier frequency. The control angle $\varphi$ allows us to choose between x- and y-pulses:
After moving to the rotating frame of the electron $\omega_e$, setting $\omega_c=  \omega_e$ and applying the RWA, the control Hamiltonian  becomes 
\begin{equation}
    H_c = \underbrace{\Omega(t) \cos(\varphi)}_{\Omega_x} S_x  + \underbrace{\Omega(t)\sin(\varphi)}_{\Omega_y} S_y,
\end{equation}
while the static part of the Hamiltonian transforms to
\begin{equation}\label{eq:Hstatic}
    H_{\text{static}} = -\omega_I I_z + A_{zz} S_z I_z + A_{zx} S_z I_x.
\end{equation}
In the product basis $\{\ket{\uparrow_e\uparrow_n}, \ket{\uparrow_e\downarrow_n}, \ket{\downarrow_e \uparrow_n}, \ket{\downarrow_e\downarrow_n} \}$, the Hamiltonian $H_\text{static}$ takes the form
\begin{equation}
    \label{eq:blockdiagonal}
    H_\text{static} = \begin{pmatrix}
        a & b & 0 & 0 \\
        b & -a & 0 & 0 \\
        0 & 0 & c & -b \\
        0 & 0 & -b & -c
    \end{pmatrix},
\end{equation}
with
\begin{equation}
    \label{eq:abc-definitions}
    a = \tfrac{1}{4}(A_{zz} - 2\omega_I), \quad 
    b = \tfrac{1}{4} A_{zx}, \quad
    c = -\tfrac{1}{4}(A_{zz} + 2\omega_I).
\end{equation}
This form is block-diagonal, reflecting the fact that the nuclear spin Hamiltonian depends on the electron spin state ($\ket{\uparrow_e}$ or $\ket{\downarrow_e}$).
\\
For the electron $\ket{\uparrow_e}$ subspace, the eigenenergies are 
\begin{equation}
    E^{\pm}_{\uparrow_e} = \pm \sqrt{a^2+b^2},
\end{equation}
with eigenvectors
\begin{equation}
    v_{\uparrow_e}^{(+)} = 
    \begin{pmatrix}
        \cos(\Theta_{\uparrow_e}/2) \\[4pt] \sin(\Theta_{\uparrow_e}/2)
    \end{pmatrix}, \quad
    v_{\uparrow_e}^{(-)} = 
    \begin{pmatrix}
        -\sin(\Theta_{\uparrow_e}/2) \\[4pt] \cos(\Theta_{\uparrow_e}/2)
    \end{pmatrix},
\end{equation}
where $\Theta_{\uparrow,e} = \text{Arctan2}(a,b) $ is obtained via a standard $2 \times 2$ diagonalization.
\\
For the $\ket{\downarrow_e}$ subspace, we obtain similarly
\begin{equation}
    E^{\pm}_{\downarrow_e} = \pm \sqrt{c^2+b^2},
\end{equation}
with corresponding eigenvectors
\begin{equation}
    v_{\downarrow_e}^{(+)} = \begin{pmatrix}
        \cos(\Theta_{\downarrow_e}/2) \\ \sin(\Theta_{\downarrow_e}/2)
    \end{pmatrix}
    , \quad
    v_{\downarrow_e}^{(-)} = \begin{pmatrix}
        -\sin(\Theta_{\downarrow_e}/2) \\ \cos(\Theta_{\downarrow_e}/2)
    \end{pmatrix}
\end{equation}
and $ \Theta_{\downarrow,e} =  \text{Arctan2}(c,-b) $.
Thus, in the joint eigenbasis of the nuclear-spin Hamiltonian, the diagonalized static Hamiltonian reads
\begin{equation}
    \tilde{H} = V^\dagger H V 
    = \mathrm{diag}(E^+_{\uparrow_e}, E^-_{\uparrow_e}, E^+_{\downarrow_e}, E^-_{\downarrow_e})
\end{equation}
where 
\begin{equation}
    V = \begin{psmallmatrix*}[r]
        \cos(\Theta_{\uparrow_e}/2) & -\sin(\Theta_{\uparrow_e}/2) & 0 &0 \\
        \sin(\Theta_{\uparrow_e}/2) & \cos(\Theta_{\uparrow_e}/2) & 0 & 0\\
        0 & 0 &\cos(\Theta_{\downarrow_e}/2)& -\sin(\Theta_{\downarrow_e}/2)\\
        0 & 0 &\sin(\Theta_{\downarrow_e}/2) & \cos(\Theta_{\downarrow_e}/2)
    \end{psmallmatrix*}
\end{equation}
is the matrix diagonalizing $H$ in the shown way.
\\
In \autoref{fig:pauli_decomp_ham} we show how the relative strength of the interaction relates to the eigenvalues depending on the Larmor frequency $\omega_I$.
Transforming the control term (with $\Omega_y = 0$) yields
\begin{equation}
    V^\dagger \Omega S_x V = \Omega \qty[ \cos(\vartheta/2) S_x - 2 \sin(\vartheta/2) S_y I_y ]
    \label{eq:eigenframe}
\end{equation}
where $\vartheta = \Theta_{\uparrow_e}-\Theta_{\downarrow_e}$.
The first term corresponds to single-electron spin flips, while the second term corresponds to double-quantum transitions (DQT), which simultaneously flip both electron and nuclear spins (discussion can be found in~\cite{patel2024single, schweiger2001principles, o2025individual, travesedo2025all}, while DQT gates for two coupled NVs are demonstrated in~\cite{morillasrozas2025entanglementgenerationdoublequantum}).
These results can be interpreted as an exact version of a Schrieffer-Wolff (SW) transformation~\cite{Schrieffer1966}, where the Hamiltonian is partitioned into an unperturbed diagonal part $H_0$ and a transverse perturbation $H_p$, such that: 
\begin{align}
    H_\text{static} &= H_0 + H_p\\
    H_0&= \text{diag}(a,-a,c,-c)\\
    H_p &= A_{zx} S_z I_x
\end{align}
The generator of the SW transformation, $G$ (defined via $[H_0, G] = H_p $),  is given by
\begin{align}
    \label{eq:G_SzIx}
    G= -\frac{i b(a-c)}{2ac} I_y + \frac{i b(a+c)}{ac} S_zI_y.
\end{align}
Applying this transformation with $V\approx e^{-G}$ to the drive term yields an approximation of the effective drive Hamiltonian from Eq.~\eqref{eq:eigenframe}
\begin{align}
    H_d' = V^\dagger \Omega S_x V  \approx  \Omega \qty[ \cos(\alpha) S_x - 2 \sin(\alpha) S_y I_y ],
\end{align}
where the rotation angle is defined as $\alpha = \frac{b(a+c)}{2ac}$.\\
In the high field limit $|\omega_I| \gg |A_{zz}|, |A_{zx}|$, the electron-nuclear mixing is weak, suppressing the contribution of DQT. 
Expanding to leading order (see \autoref{sec:appendix}) gives
\begin{equation}\label{eq:DQT-ratio}
    \sin(\vartheta/2) \approx -\frac{A_{zx}}{2\omega_I}\approx \sin\alpha,
\end{equation}
such that the effective entangling interaction strength is proportional to $A_{zx}$ and the two approaches give the same result to first order. 
This result highlights a key limitation: while transverse hyperfine coupling enables entanglement via DQT processes, the rate is strongly suppressed at large nuclear Zeeman splittings. 
In this paper, we make use of the strong coupling regime and the tunability of $\omega_I$ to realize different quantum gates and investigate their speed limits.
\subsection{Optimal Microwave Control}
\label{sec:optimal_control}
One method for realizing effective control,  includes using quantum optimal control, which is formulating the optimization task in a framework of an objective-function minimization evaluating the difference between realized and wanted dynamics~\cite{brif2010control, palao2013steering, glaser2015training, barnes2017fast, rembold2020, preti2022continuous, muller2022one,calzavara2025classical} by direct search for experimentally feasible MW pulse shapes that implement the desired two-qubit gates within the available hardware constraints.
Here, we employ the dressed Chopped RAndom Basis (dCRAB)~\cite{caneva2011chopped, doria2011optimal, rach2015dressing, muller2022one} algorithm.
We combine this algorithm with Numba-compiled numerical evaluation improving efficiency of evaluating the objective function.
A basis expansion allows to reduce the optimization problem to a few-parameter optimization with built-in experimental constraints, e.g. that the hardware does not need to do hard switches between largely varying amplitudes.
The dCRAB algorithm expands the envelope functions $\Omega_k(t)$, $k \in \{X,Y\}$ in a basis $\{f_i\}_{i=1,\cdots, N_c}$ such that the total envelope takes the form 
\begin{equation}
    \Omega_k(t)= E(t) \sum_{i=1}^{N_c} c_i f_i(t)
\end{equation}
where $c_i$ are the coefficients of the respective basis functions and $E(t)$ is some general envelop function which can be used to enforce the pulse to start and end at zero amplitude~\cite{muller2022one}.
In our case we used
\begin{equation}
E(t)=
\begin{cases}
\alpha\!\left[
e^{-\frac{(t-t_r)^2}{2\sigma^2}}
-
e^{-\frac{t_r^2}{2\sigma^2}}
\right], & 0<t\le t_r, \\[0.8em]

1, & t_r<t<t_f-t_r, \\[0.8em]

\alpha\!\left[
e^{-\frac{(t-(t_f-t_r))^2}{2\sigma^2}}
-
e^{-\frac{t_r^2}{2\sigma^2}}
\right], & t_f-t_r\le t<t_f .
\end{cases}
\end{equation}
with $\alpha = \left[1-\exp\!\left(-\frac{t_r^2}{2\sigma^2}\right)\right]^{-1}$
being a normalization constant, $t_r=\SI{100}{\nano \s}$ denotes the rise time, and
$\sigma = t_r/4$ sets the width of the Gaussian ramps.

The optimization of the parameters $c_i$ can now be done by choosing a suitable optimization algorithm.
However, a fixed choice of $N_c$ (here $N_c=20$) may introduce local minima in the control landscape and restrict the accessible control space~\cite{muller2022one}.
This can be addressed by introducing super-iterations $j=1,\cdots ,N$ (here N=10) such that in the $j$-th super iteration we optimize the coefficients $c_i^{j}$ of
\begin{equation}
        f^j(t) = c_0^j f^{j-1}(t) + \sum_{i=1}^{N_c} c_i^j f_i^j(t)
\end{equation}
where $f^{j-1}(t)$ is the resulting function of the optimization of the previous super iterations, $c_0^j$ its weight and $f_i^j(t)$ are new basis functions (specifically we chose the Fourier basis where the number of oscillations allowed during the pulse time was randomly chosen between $0.1$ and $40$) which are chosen such that they do not have a great overlap in the search space~\cite{muller2022one}.

As an objective function which we want to minimize, we define a figure of merit based on the Frobenius norm as the gate infidelity
\begin{equation}
    1 - \mathcal{F} = 1 - \frac{1}{d^2} \abs{\Tr{W U_{T}^\dagger}}^2\,,
    \label{eq:unitfid}
\end{equation}
where $W$ denotes the simulated time-evolution operator, $U_T$ is the target unitary and $d$ is the dimension of the Hilbert space (for the two-qubit gates $d=4$).
Instead, if we are only interested in the non-local part of this two-qubit gate operation we use~\cite{watts2015optimizing,Goerz2015, frey2026optimaltwoqubitgatesgroupiv}
\begin{equation}
        \label{eq:weylfid}
        1-\mathcal{F}_{\text{nl}}(V_G, U_{\text{nl}}) = 1-\cos(\frac{\Delta c_1}{2})\cos(\frac{\Delta c_2}{2})\cos(\frac{\Delta c_3}{2}).
\end{equation}
Here, $\Delta c_i = c_i^{(G)} - c_i$ are the Weyl coordinates differences, and we call Eq.~\eqref{eq:weylfid} the Weyl-induced figure of merit.

One can check that the system is fully controllable and if no constraints are given to the control functions the QSL for different quantum gates can then in principle be obtained by optimizing the controls for different gate times $T$ and identification of a convergence region for the infidelity~\cite{Caneva2009}.
In the next few sections we will discuss how the gate time limit can be found out in a rigorous, analytical way, dependent on the strategy used. 

\section{Two-Qubit Protocols and Results}
\label{sec:Two-Qubit-Protocols}
\subsection{Dynamical decoupling for ZX and ZZ Gates}
\begin{figure}
    \centering
    \includegraphics{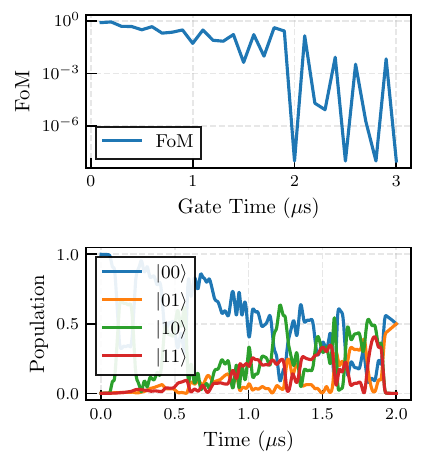}
    \caption{Optimization of a ZX/2 gate with the dCRAB optimal control method. The top plot shows the resulting FoM with control pulses which were amplitude-limited to $\abs{\Omega_{\text{max}}} = 2\pi\times\SI{15}{\MHz}$ to ensure experimental feasibility. The pulse was multiplied by a Gaussian envelope function ensuring that it starts and ends at zero power (see \autoref{sec:optimal_control}). The lower plot shows the population of the two-qubit system during the applied pulses for a gate time of $T=\SI{2}{\mu s}$. The gate reached a gate fidelity of $1-\mathcal{F} < 10^{-8}$, while we set the Larmor frequency to $\omega_I = 2 \pi \times \SI{3}{\MHz}$.}
    \label{fig:ZX-optimized}
\end{figure}

We start our analysis of gate protocols with dynamical decoupling (DD) sequences applied to the electron spin.
We move into the interaction frame of the nuclear spin, where the Hamiltonian becomes
\begin{align}
    \Tilde{H}(t) &= e^{i \omega_I t I_z} H e^{-i\omega_I t I_z} - \omega_I I_z \\
              &= A_{zz} S_z I_z + A_{zx} \qty(\cos(\omega_I t) S_z I_x - \sin(\omega_I t) S_z I_y). \nonumber
\end{align}
In this form, the transverse hyperfine term oscillates with the nuclear Larmor frequency, suggesting that properly timed control pulses can selectively average it into an effective coupling~\cite{meiboom1958modified}.
However, if we let the system evolve freely, the oscillating terms will average out over a full period and therefore approximately implement a ZZ/2 $=\exp(-i \pi S_z I_z)$  gate. The conditions for this gate are 
\begin{equation}
    \omega_I = 2 A_{zz}= 2\pi \times \SI{5.72468}{\MHz} ,\quad T = \frac{\pi}{A_{zz}}\approx \SI{175}{ns}.  
\end{equation}
For the implementation of the entangling gate
\begin{equation}
    \text{ZX/2} = e^{-i \frac{\pi}{4} \sigma_z \otimes \sigma_x} = e^{-i \pi S_z I_x},
\end{equation}
we use average Hamiltonian theory~\cite{haeberlen1968coherent,Vandersypen2005}. The Hamiltonian between DD pulses in an interval $\delta_j \leq t \leq \delta_{j+1}$ is
\begin{align}
    H_{\text{avg}}(\delta_{j+1}, \delta_j) &= \frac{1}{\delta_{j+1}- \delta_j} \int_{\delta_j}^{\delta_{j+1}} H(t) dt \nonumber \\
                   &= A_{zz} S_z I_z +\frac{A_{zx} S_z}{\omega_t (\delta_{j+1}-\delta_j)}\\
           & [\qty(\sin(\omega_I \delta_{j+1})-\sin(\omega_I \delta_{j})) I_x \nonumber \\
           &+\qty(\cos(\omega_I \delta_{j+1})- \cos(\omega_I \delta_j))I_y]. \nonumber
\end{align}
This can be taken further by taking a $n$-pulse Carr-Purcell-Meiboom-Gill-Sequenz (CPMG)~\cite{Carr1954} sequence with pulse locations $\delta_j = T \tfrac{j-1/2}{n}$, $\delta_0 = 0$, and $\delta_{n+1} = T$. 
In this case, the toggling-frame Hamiltonians add up to
\begin{equation}
    H_{\text{tot}} = \sum_{j=0}^n (-1)^j H_{\text{avg}}(\delta_j, \delta_{j+1}).
\end{equation}
If the nuclear Larmor frequency is tuned such that $\frac{T \omega_I}{n}\overset{!}{=} \pi$ and $n$ is even for canceling unwanted terms, this expression simplifies to
\begin{equation}
    \int_0^T H(t)\, dt \;\approx\; \frac{2}{\pi} A_{zx} T S_z I_x \;\overset{!}{=}\; \pi S_z I_x.
\end{equation}
Thus, the required Larmor frequency and gate duration are
\begin{align}
    \omega_I &= \frac{2 n A_{zx}}{\pi}, 
    \qquad T = \frac{\pi^2}{2 A_{zx}}\approx \SI{1.303}{\mu s},
\end{align}
showing that the achievable gate speed with decoupling gates is fundamentally limited by the transverse coupling $A_{zx}$.
\begin{figure}
    \centering
    \includegraphics{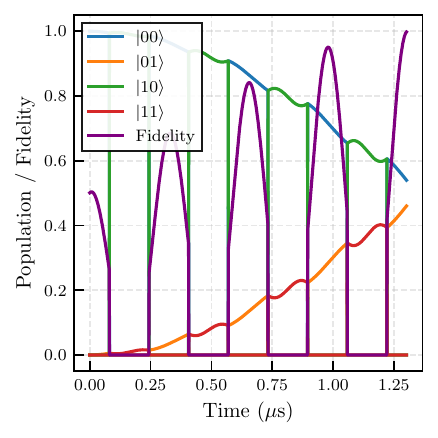}
    \caption{Population dynamics during a dynamical decoupling gate implemented with $n=8$ instantaneous $\pi$-pulses and a Larmor frequency of $\omega_I = \frac{16 A_{zx}}{\pi} \approx 2 \pi \times \SI{3.070}{\MHz}$. 
    The gate realizes the entangling ZX/2 operation. 
    The purple line shows the unitary fidelity Eq.~\eqref{eq:unitfid}, which reaches $\mathcal{F} = 99.87\%$ for only eight pulses, and can be systematically improved by increasing $n$.}
    \label{fig:DDgate}
\end{figure}
In \autoref{fig:DDgate} we show the time evolution of this technique for $n=8$.
In this case we have to choose $\omega_I=2\pi\times\SI{3.07}{MHz}$. Note that we use the Lamor frequency to tune the interaction strength (or rotation angle of the gate). Since this is a single parameter, this simple strategy would not work for more than one nuclear spin. However, slightly more complex pulse sequences (where each pulse is substituted by a composite pulse) allow to tune the rotation angle of the gate directly~\cite{Casanova2015}.
In practice, however, the usefulness of this approach is restricted by experimental limitations: high-fidelity DD gates require rapid and precise $\pi$-pulses on the electron spin, which are challenging to realize with available MW control~\cite{grimm2025coherent, resch2025high, klotz2022prolonged}.
As a result, while a powerful tool for weakly-coupled spin systems~\cite{Casanova2015,beukers2025control,takou2024generation}, DD-based entangling gates are unlikely to be the most practical route for strongly-coupled group-\uproman{4} systems.
Indeed, if we consider the pulse-amplitude limitation $\abs{\Omega_{\text{max}}} = 2\pi\times\SI{15}{\MHz}$, the duration of the 8 $\pi$-pulse time is already $\approx \SI{0.27}{\micro s}$, which is a substantial part of the gate time. If we nevertheless want a high-fidelity gate, we have to increase slightly the gate time and employ QOC to optimize the envelope of the MW control pulse.
\autoref{fig:ZX-optimized} shows the achieved FoM   $1-\mathcal{F}$ (top panel) for an optimization of the ZX/2 gate with dCRAB as a function of the gate time. The lower panel shows the corresponding population dynamics of the two-qubit system during the application of the optimized pulse for $T=\SI{2}{\mu s}$. 
The optimized pulses achieve a gate error of $1-\mathcal{F} < 10^{-8}\%$, showing that the use of QOC allows to take into account the amplitude constraint and lead to an increase of the fidelity.
%
\subsection{Double-Quantum-Transition (YY) Gate}
\begin{figure}
    \centering
    \includegraphics{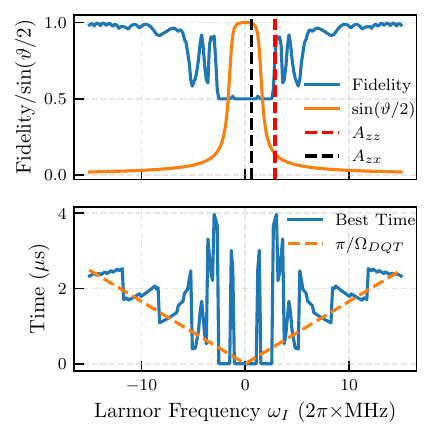}
    \caption{
    Performance of the double-quantum transition (DQT) gate as a function of the nuclear Larmor frequency $\omega_I$. 
    Top panel: maximum gate fidelity obtained using rectangular MW pulses with a Rabi frequency of $\Omega = 2\pi \times \SI{10}{\MHz}$, while varying the gate duration between $0$ and $\SI{4}{\us}$. 
    Bottom panel: corresponding gate times at which the optimal fidelity was achieved. 
    The orange line indicates a theoretical speed estimation derived in ~\autoref{sec:background}, showing good agreement with the observed $1/{\omega_I}$ scaling  of Eq.~\eqref{eq:Omega_DQT}.}
    \label{fig:fids_vs_lamor}
\end{figure}

\begin{figure}[t]
    \centering
    \includegraphics{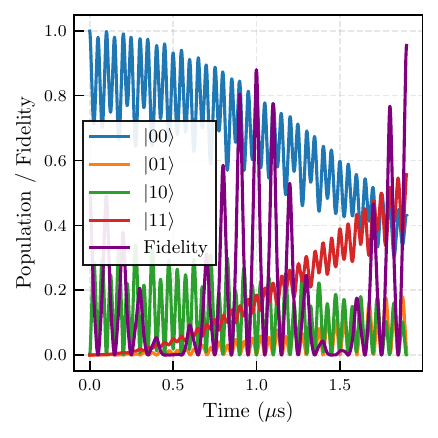}
    \caption{
    Time evolution of a two-qubit system under DQT driving, shown in the rotating frame of the electron spin after applying the RWA. 
    The control field is of the form $\Omega(t) = \Omega_0 \cos(\omega_I t)\cos(\omega_e t)$, with $\Omega_0 = \SI{11}{\MHz}$ and a gate time of $T=\SI{1.9}{\micro \s}$ , leading to coherent population oscillations between $|00\rangle=\ket{\uparrow_e\uparrow_n}$ and $|11\rangle=\ket{\downarrow_e\downarrow_n}$ mediated by the $A_{zx}$ coupling. In this plot we set the Larmor frequency to $\omega_I = 2\pi \times \SI{10}{\MHz}$.}
    \label{fig:DQT_pops}
\end{figure}

\begin{figure}
    \centering
    \includegraphics{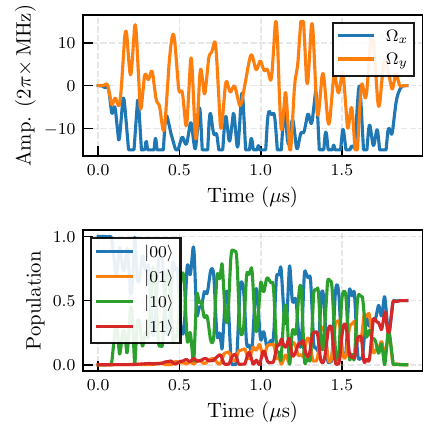}
    \caption{Shown is a dCRAB Optimization of a YY/2 gate with the optimal control method. The top plot shows the resulting optimal control pulses which were amplitude-limited to $\abs{\Omega_{\text{max}}} = 2\pi\times\SI{15}{\MHz}$, the start was a DQT gate with smooth envelope and a pure X-drive. The lower plot shows the population of the two-qubit system. The final gate performance is $1-\mathcal{F} = \SI{6.07e-7}{}$. The Larmor frequency was included in the optimization process and ended up to be $\omega_I = 2 \pi \times \SI{11.045}{\MHz}$.}
    \label{fig:optimized-YY}
\end{figure}

Having established how the transverse hyperfine term $A_{zx}$ enables otherwise forbidden flip-flop and flip-flip processes via the effective $S_yI_y$ (or $\sigma_y\otimes\sigma_y$) control term of Eq.~\eqref{eq:eigenframe}, we now examine how this interaction can be harnessed to implement two-qubit gates. 
These simultaneous flips of the electron and nuclear spins (or DQT), mediated by the transverse hyperfine term $A_{zx}$ allow to generate another universal entangling gate for the GeV system. 
In the eigenbasis of $H_{\text{static}}$, the control Hamiltonian enables coherent coupling between the states $\ket{\uparrow\uparrow}$ and $\ket{\downarrow\downarrow}$. 
For large enough nuclear Larmor frequencies $\omega_I$ (see \autoref{eq:DQT-ratio}), the effective Rabi frequency of the DQT channel is, assuming only $S_x$ driving (i.e. $\Omega_y(t) = 0$), approximately
\begin{equation}
    \Omega_{\mathrm{DQT}} \approx \frac{A_{zx}}{\omega_I}\,\Omega,
    \label{eq:Omega_DQT}
\end{equation}
driving $\Omega_{\text{DQT}} S_y I_y$, where the sign of $\Omega$ is chosen accordingly.
The corresponding two-qubit entangling gate
\begin{equation}
    U = \exp\!\left(-i \tfrac{\pi}{4}\,\sigma_y \otimes \sigma_y\right)
\end{equation}
can thus be approximately achieved in a gate time of
\begin{equation}
    \label{eq:scaling}
    T \approx \frac{\pi}{\abs{\Omega_{\mathrm{DQT}}}}.
\end{equation}
However, the drive has to be weak enough to drive the transitions selectively. A lower bound on the time required to implement the target operation is given by
\begin{equation}
    T \geq \frac{\pi}{\sqrt{A_{zx}^2 + A_{zz}^2}} \approx \SI{171}{\nano\second},
\end{equation}
derived using results from~\cite{basyildiz2023speed} and~\cite{impens2025approaching} under the assumption of full single-qubit control  over both the electron and nuclear spin.
It allows the combined hyperfine interaction to be rotated to the YY direction,
thereby achieving an efficient evolution.
More details on the achievable gate times for different target gates can be found in \autoref{sec:algebraic}. 
\\
In our case, however, we restrict ourselves to limited power MW control on the electron spin, and no (rf)-control on the nuclear spin
as fast nuclear spin manipulation via rf pulses is experimentally unfeasible.
For our hyperfine parameters, constant-amplitude (rectangular) MW pulses with $\Omega = 2\pi\times \SI{10}{\MHz}$ produce gate durations of a few microseconds, which is still below coherence times achievable under current experimental conditions~\cite{senkalla2024germanium, banasiak2026diamond}.
In \autoref{fig:fids_vs_lamor} we illustrate this behavior. For each value of the Lamor frequency we determine the gate time (out of the interval $\SIrange{0}{4}{\micro s}$) that maximized the fidelity. This required gate time increases with increasing (absolute value of the) nuclear Larmor frequency, consistent with the scaling predicted by Eq.~\eqref{eq:Omega_DQT} and Eq.~\eqref{eq:scaling}. In this way, the maximum fidelity as a function of the nuclear Lamor frequency is achieved for the largest investigated values of the Lamor frequency.
The maximum of $F=99.6\%$ was reached for $\omega_I = \pm 2 \pi \times \SI{13.65}{\MHz}$.
For small values of $\omega_I$ we find worse fidelity and especially around $\pm A_{zz}$ also the gate time increases. This can be understood by considering \autoref{fig:pauli_decomp_ham} where we see the relative contributions of the different interaction terms (that determine the gate time) and the eigenvalues of the Hamiltonian that determine how well we can resolve the two different drive terms (and thus determine the fidelity of the gate).
The corresponding dynamics are illustrated in ~\autoref{fig:DQT_pops}, which shows coherent oscillations between $\ket{\uparrow_e\uparrow_n}$ and $\ket{\downarrow_e\downarrow_n}$ driven by the DQT mechanism.
Furthermore, we also did a dCRAB optimization of the DQT YY-gate.
First, we smoothed the envelope of applied pulses by a Gaussian raise time of $\SI{100}{\nano \s}$.
We limited both $\Omega$ and $\omega_I$ to $2\pi\times \SI{15}{\MHz}$ reflecting experimental constraints, and pre-optimized these two values for a flat-top pulse. 
Afterward, we optimized this initial guess envelope with dCRAB, for reaching better fidelity. 
The results are depicted in  \autoref{fig:optimized-YY}. 
Here, the Larmor frequency $\omega_I$ was also included in the optimization process, since it is important for timing the resonance peak of the fidelity. 
\subsection{Approaching the Speed-Limit through Algebraic Decomposition}
\label{sec:algebraic}
\begin{figure}
    \centering
    \includegraphics{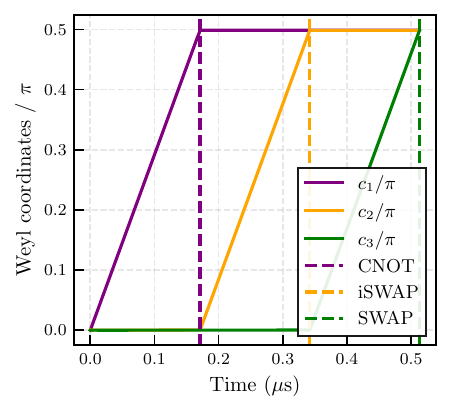}
    \caption{Time evolution of the Weyl coordinates. The system undergoes free evolution interleaved with instantaneous local transformations at $\tau^{\text{CNOT}}$ and $2\tau^{\text{CNOT}}$; these points are indicated by the dashed vertical lines. The dashed vertical lines also correspond to realizations of CNOT, iSWAP and SWAP gates, respectively.}
    \label{fig:algebraicdecomposition}
\end{figure}

In this section, we employ the Cartan (or KAK) decomposition~\cite{Khaneja2000, Zhang2003b,Zhang2004a,Nakajima2003}
(see also~\ref{sec:localgates}) of a two-qubit gate to demonstrate the ultimate speed-limit for entangling gates in our system. Any two-qubit gate $U$ can be decomposed in its non-local content $A$ and two local gates $K_1 = k_1 \otimes l_1$ and $K_2 = k_2 \otimes l_2$, where $k_1, k_2, l_1, l_2 \in SU(2)$ such that
\begin{equation}
U = K_1 A K_2,
\end{equation}
and the non-local content A is given by
\begin{equation}
A = \exp[\frac{i}{2}
\left(c_1\, \sigma_x\otimes\sigma_x + c_2\, \sigma_y\otimes\sigma_y + c_3\, \sigma_z\otimes\sigma_z\right)],
\end{equation}
with Weyl coordinates $c_1$, $c_2$ and $c_3$.\\
If we let our system evolve freely according to Eq.~\eqref{eq:Hstatic} for some time $t$, this leads to a non-local gate content
\begin{eqnarray}
    A=\exp{i \omega_{\text{ent}} t\sigma_x\otimes\sigma_x}\,,
\end{eqnarray}
where $\omega_{\text{ent}}=\frac{\sqrt{A_{zx}^2 + A_{zz}^2}}{4}$.
Under the assumption of infinitely fast single qubit operations we can implement the CNOT ($c_1=\pi/2$, $c_2=c_3=0$), iSWAP ($c_1=c_2=\pi/2$, $c_3=0$) and SWAP ($c_1=c_2=c_3 = \pi/2$) operations in:
\begin{align}
    \tau^{\text{CNOT}} &= \frac{\pi}{\sqrt{A_{zx}^2 + A_{zz}^2}} \\
    \tau^{\text{iSWAP}}&=  2\tau^{\text{CNOT}} \\
    \tau^{\text{SWAP}} &=  3\tau^{\text{CNOT}}\,.
\end{align}
We note that for iSWAP this is the QSL calculated in~\cite{basyildiz2023speed}. In \autoref{sec:localgates} we show how to calculate the local gates that have to be applied before the free evolution and after each time interval $\tau^{\text{CNOT}}$ to realize the respective two-qubit gate (CNOT, iSWAP and SWAP). 
\autoref{fig:algebraicdecomposition} shows the time evolution of the Weyl coordinates when the free evolution is interleaved with these instantaneous gates.
Interestingly, if the system is undriven one of the Weyl coordinates builds up linearly in time. 
At the multiples of the calculated QSL $\tau^{\text{CNOT}}$ we applied the respective local gates (as instantaneous perfect rotations) and therefore enable the build-up of the next Weyl coordinate.
Furthermore, at these times (after the local gates) we find perfect realizations of the three respective gates.
The gates discussed in the previous subsections, i.e., the ZZ, ZX and YY gate, are locally equivalent to CNOT. We thus discuss here to more detail the SWAP gate (and locally equivalent gates) as the most challenging task that involves all three Weyl coefficients.
\begin{figure*}
    \centering
    \includegraphics{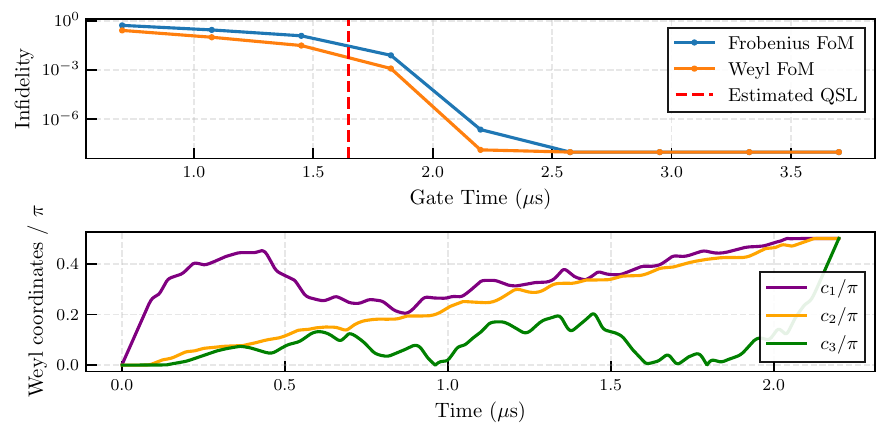}
    \caption{The upper panel shows the optimization procedure of a SWAP gate. The control of the system is optimized once with the standard Frobenius FoM \eqref{eq:unitfid} (blue line) and once with the Weyl coordinate infidelity \eqref{eq:weylfid}. The optimization was stopped when the FoM reached FoM  $\leq \SI{1e-8}{}$. The lower panel shows the time evolution of the Weyl coordinates for a gate duration of $T=\SI{2.2}{\micro \s}$. In this figure the Larmor frequency was set to $\omega_I = 2 \pi \times \SI{3}{\MHz}$.} 
    \label{fig:SWAP}
\end{figure*}
In the absence of direct single-qubit control of the nuclear spins (since the RF pulses on the nuclear spin are much slower than MW controls on the electron) the SWAP gate must be synthesized via the available MW-driven interactions.
Specifically, we have previously demonstrated the generation of ZX, YY and ZZ gates, where the ZX gate can be converted into an XX gate via local MW pulses on the electron:
\begin{equation}
    e^{-i \pi S_x I_x} = \qty(H \otimes \mathbb{1}) e^{-i \pi S_z I_x} \qty(H \otimes \mathbb{1}).
\end{equation}
The total duration for such a sequence is approximately the sum of the individual gate times, resulting in $T \approx \SI{1.649}{\micro \s}$ (not counting the MW-driven Hadamards). 
The actual gate time may be slightly longer due to the maximum pulse amplitude constraint of $\Omega_{\text{max}} = 2 \pi \times \SI{15}{\MHz}$.
These results are confirmed in \autoref{fig:SWAP}, where the upper panel shows the convergence of the respective figures of merit.
The lower panel illustrates the temporal evolution of the Weyl coordinates for a gate duration of $T= \SI{2.2}{\micro \s}$.
The non-linear evolution of the Weyl coordinates indicates that the QSL for this operation, under the given constraints, has not yet been reached.
While the theoretical time scale is $T = \SI{1.6}{\micro \s}$, the observed convergence region of the FoMs around $T = \SIrange{2}{2.2}{\micro \s}$ for the FoM based on the Weyl coordinated (SWAP equivalent) and slightly higher for the FoM based on the Frobenius norm (SWAP gate).

We see that a fast, high fidelity SWAP gate can already be implemented in a reasonably short time as confirmed also for noisy evolution in Ref.~\cite{frey2026optimaltwoqubitgatesgroupiv}. 

%
\section{Discussion}
\label{sec:discussion}
The four approaches considered in this work, i.e., (DQT) gates, DD gates,  MW-driven QOC, and free evolution with interleaved fast single-qubit rotations each highlight different pathways toward implementing fast (compared to RF control) and high-fidelity two-qubit gates in GeV centers. 
\\
DQT gates directly exploit the transverse hyperfine interaction $A_{zx}$, enabling simultaneous electron–nuclear spin flips. 
While conceptually simple, their performance is intrinsically limited by the smallness of $A_{zx}$ and the need to resolve the DQT spectrally, which constrains both gate speed and robustness at large nuclear Larmor frequencies (see \autoref{fig:pauli_decomp_ham}). 
As a result, DQT gates are attractive for proof-of-principle demonstrations but less promising in practice.
However, they might offer the possibility for a guess in optimal control setups.
DD gates, in contrast, provide a systematic way to engineer effective interactions through pulse sequences. 
We showed that entangling $ZX/2$ gates with fidelities exceeding 99.8\% can be obtained with only a modest number of pulses in idealized simulations. 
However, this method relies on the rapid application of high-fidelity $\pi$-pulses, which remains a major experimental challenge for strongly-coupled nuclear spins given the finite MW power. 
Thus, while DD gates are powerful in weakly coupled systems~\cite{Casanova2015,beukers2025control,takou2024generation}, their practical implementation in strongly-coupled group-IV centers is not feasible with current experimental setups.
Given the experimental constraints, a method that does not require idealized, instantaneous $\pi$-pulses is given via the use of MW-driven QOC. 
It provides a complementary approach that avoids both of these limitations.
By tailoring smooth control fields subject to realistic amplitude constraints, we demonstrated that entangling gates can be realized with very high fidelity. 
Importantly, QOC makes no explicit assumption about the relative strength of couplings or the need for ideal $\pi$-pulses, and can be readily extended to incorporate robustness against noise or drift in experimental parameters as has been demonstrated in~\cite{frey2026optimaltwoqubitgatesgroupiv}. 
This flexibility suggests that QOC is the most practical route toward two-qubit operations for group-\uproman{4} systems with strongly-coupled nuclear spins. Indeed, recent QOC results for NV centers with a few nuclear spins are based on an approach that could also be adapted to group-\uproman{4} systems~\cite{wang2025,baran2026}.
%
\section{Conclusion}
\label{sec:conclusion}
Our results highlight both the opportunities and the challenges of realizing two-qubit gates in group-\uproman{4} vacancy centers. 
We have shown that, careful exploitation of transverse hyperfine coupling and the use of advanced control techniques such as dynamical decoupling gates and optimal MW control pulse shaping can enable high fidelity two-qubit operations.
These approaches bring gate fidelities closer to the threshold required for distributed quantum information processing, providing a potential path toward scalable quantum architectures. 
Nevertheless, the sensitivity of gate performance to experimental imperfections and to decoherence due to the environment needs further refinement of control protocols and error mitigation strategies.
Future work should explore the integration of these optimized control schemes into existing quantum protocols and investigate adaptive feedback with experimental benchmarking of these techniques in state-of-the-art GeV devices. Furthermore, the QOC protocols should be extended to more than one nuclear spin.
Looking forward, the ability to implement entangling gates in GeV centers positions them as viable candidates as building blocks for quantum networks and distributed quantum computing. 
Beyond GeV centers, the methods developed here can also be used as strategies for other group-IV vacancy systems, solid-state spin platforms and scalable quantum technologies~\cite{grimm2025coherent, muller2022one, egger2014adaptive,ruf2021quantum,o2025individual}.
\section{Acknowledgements}
Wie thank P. Vetter, K. Senkalla, S.G. Walliser and P. Gundlapalli for insightful discussions.
This work was supported by Germany’s Excellence Strategy – Cluster of Excellence Matter and Light for Quantum Computing (ML4Q) EXC 2004/1 – 390534769,  the European Union’s HORIZON Europe program via projects SPINUS (No. 101135699) and OpenSuperQPlus100 (No. 101113946), by AIDAS-AI, Data Analytics and Scalable Simulation, which is a Joint Virtual Laboratory gathering the Forschungszentrum J\"ulich and the French Alternative Energies and Atomic Energy Commission, as well as by BMFTR via the project SPINNING (No. 13N16210).

\section*{Data Availability}
The data presented in this study is available under \url{https://doi.org/10.5281/zenodo.19660411
}.

\section{Author contributions}
The simulation and analysis was done by J.F. and M.M.M.; F.W. and M.M.M. supervised the project. All authors discussed the results. 

\appendix
\section{Approximation of the Eigen difference angle }
\label{sec:appendix}
To obtain an explicit analytic expansion of the eigen angle representation in Eq.~\eqref{eq:eigenframe} we use the standard trigonometric identities
\begin{equation}
    \sin(\arctan x) = \frac{x}{\sqrt{1+x^2}}, \quad 
    \cos(\arctan y) = \frac{1}{\sqrt{1+y^2}},
\end{equation}
where
\begin{equation}
    x = \frac{\sqrt{a^2+b^2}-a}{b}, \quad
    y = \frac{c - \sqrt{c^2+b^2}}{b}
\end{equation}
and $a,b,c$ are defined in Eq.~\eqref{eq:abc-definitions}.
Applying these substitutions to the eigen rotation angle difference, we find
\begin{align}
    \sin\left(\frac{\vartheta}{2}\right) 
    &= \sin\left(\frac{\Theta_{\uparrow_e}}{2}\right)\cos\left(\frac{\Theta_{\downarrow_e}}{2}\right) 
       - \cos\left(\frac{\Theta_{\uparrow_e}}{2}\right)\sin\left(\frac{\Theta_{\downarrow_e}}{2}\right) \nonumber \\
    &= \frac{x}{\sqrt{1+x^2}} \cdot \frac{1}{\sqrt{1+y^2}} 
       - \frac{1}{\sqrt{1+x^2}} \cdot \frac{y}{\sqrt{1+y^2}} \\
    &= \frac{x - y}{\sqrt{(1+x^2)(1+y^2)}}. \nonumber
\end{align}
Series expansion of this expression in the large Larmor frequency limit $\omega_I \gg \abs{A_{zx}}, \abs{A_{zz}}$ results in
\begin{equation}
    \sin\frac{\vartheta}{2} \approx 
        -\frac{A_{zx}}{2 \, \omega_I} 
        + \frac{A_{zx} (A_{zx}^2 - 2 A_{zz}^2)}{16 \, \omega_I^3} 
        + \mathcal{O}\!\left(\frac{1}{\omega_I^5}\right).
\end{equation}
Thus, in leading order the strength of the DQT interaction depends on the ratio of the transverse coupling term $A_{zx}$ and the Larmor frequency $\omega_I$ while higher orders also depend on the longitudinal coupling strength $A_{zz}$.
\section{Comment on Schrieffer-Wolff transformation}
To find the generator that satisfies $[H_0, G] =  H_p$ we use the relations
\begin{align}
    4 S_z^2 =\mathbb{1}
\end{align}
and 
\begin{align}
    [I_y, I_z] = i I_x
\end{align}
from these we can conclude that G should be of the form
\begin{align}
    G = \alpha I_y + \beta S_z I_y
\end{align}
Solving the condition $[H_0, G] = H_p$ for $\alpha$ and $\beta$ yields Eq.~\eqref{eq:G_SzIx}.
The SW transformation assumes the perturbation to be small and therefore corresponds to a small angle approximation. 
\section{Local corrections from the Cartan Decomposition}
\label{sec:localgates}
As shown in Eq.~\eqref{eq:blockdiagonal} the Hamiltonian of the system is in a block-diagonal form
\begin{equation}
    H = \begin{pmatrix}
        H_1 & 0\\
        0 & H_2
    \end{pmatrix},
\end{equation}
which implies that also the resulting time propagators are block diagonal 
\begin{equation}
        U = \begin{pmatrix}
        U_1 & 0\\
        0 & U_2
    \end{pmatrix}.
\end{equation}
We know that after the characteristic entangling time $\tau^{\text{CNOT}}=\frac{\pi}{\sqrt{A_{zx}^2 + A_{zz}^2}}$, the accumulated interaction equates to a CNOT equivalent that is we have (up to a global phase)
\begin{equation}
    \label{eq:kak}
    U(\tau^{\text{CNOT}}) = K_1 \text{CNOT} K_2.
\end{equation}
Now, we want to calculate the local-unitaries needed to implement a real CNOT.
We start the calculations by assuming the following form of the local gates
\begin{equation}
    K_1 = \begin{pmatrix}
        1&0\\
        0&e^{i \phi}
    \end{pmatrix} \otimes C_1, K_2 = \mathbb{1} \otimes  C_2
\end{equation}
where $C_1, C_2 \in U(2)$. 
Inserting these assumptions into Eq.~\eqref{eq:kak} we get the conditions 
\begin{align}
    U_1 = C_1 C_2, U_2 = e^{i \phi} C_1 X C_2.
\end{align}
Combining these two we have
\begin{align}
    \label{eq:Mctwo}
    M = U_1^\dagger U_2 = e^{i \phi} C_2^\dagger X C_2. 
\end{align}
Since $M \in U(2)$ it has the eigenvalues $\lambda_{\pm} = e^{\pm i \phi}$ and we can write $M$ as 
\begin{align}
    M = e^{i \phi} V Z V^\dagger.
\end{align}
If we now use the fact $X = H Z H$ we get by comparing the last equation with Eq.~\eqref{eq:Mctwo}
\begin{align}
     C_1  = U_1 V H, \quad C_2 = H V^\dagger.
\end{align}
With this result we can finally write our wanted CNOT operation as
\begin{align}
    \text{CNOT}_{12} = K_1^\dagger U K_2^\dagger.
\end{align}
The CNOT on the other qubit is writable as
\begin{align}
    \text{CNOT}_{21} = \qty(H \otimes H) \, \text{CNOT}_{12} \, \qty(H \otimes H)
\end{align}
such that we can construct a SWAP gate by
\begin{align}
    \text{SWAP} = \text{CNOT}_{12} \text{CNOT}_{21} \text{CNOT}_{12}
\end{align}
This is demonstrated in \autoref{fig:algebraicdecomposition}.

\bibliographystyle{apsrev4-2}
\bibliography{references} 

\end{document}